\documentstyle[12pt,psfig]{article}
\makeatletter
\def\@cite#1#2{{[{#1}]\if@tempswa\typeout
{IJCGA warning: optional citation argument
ignored: `#2'} \fi}}

\newcount\@tempcntc
\def\@citex[#1]#2{\if@filesw\immediate\write\@auxout{\string\citation{#2}}\fi
  \@tempcnta\z@\@tempcntb\m@ne\def\@citea{}\@cite{\@for\@citeb:=#2\do
    {\@ifundefined
       {b@\@citeb}{\@citeo\@tempcntb\m@ne\@citea\def\@citea{,}{\bf ?}\@warning
       {Citation `\@citeb' on page \thepage \space undefined}}%
    {\setbox\z@\hbox{\global\@tempcntc0\csname b@\@citeb\endcsname\relax}%
     \ifnum\@tempcntc=\z@ \@citeo\@tempcntb\m@ne
       \@citea\def\@citea{,}\hbox{\csname b@\@citeb\endcsname}%
     \else
      \advance\@tempcntb\@ne
      \ifnum\@tempcntb=\@tempcntc
      \else\advance\@tempcntb\m@ne\@citeo
      \@tempcnta\@tempcntc\@tempcntb\@tempcntc\fi\fi}}\@citeo}{#1}}
\def\@citeo{\ifnum\@tempcnta>\@tempcntb\else\@citea\def\@citea{,}%
  \ifnum\@tempcnta=\@tempcntb\the\@tempcnta\else
   {\advance\@tempcnta\@ne\ifnum\@tempcnta=\@tempcntb \else
\def\@citea{--}\fi
    \advance\@tempcnta\m@ne\the\@tempcnta\@citea\the\@tempcntb}\fi\fi}
\makeatother

\newcommand{\dr}{\mbox{\footnotesize$\overline{\rm DR}$~}}
\newcommand{\ms}{\mbox{\footnotesize$\overline{\rm MS}$~}}

\newcommand{\tilt}{\tilde{t}}
\newcommand{\sto}{\tilde{t}_1}
\newcommand{\stw}{\tilde{t}_2}

\newcommand{\gsim}{\lower.7ex\hbox{$\;\stackrel{\textstyle>}{\sim}\;$}}
\newcommand{\lsim}{\lower.7ex\hbox{$\;\stackrel{\textstyle<}{\sim}\;$}}
\newcommand{\be}{\begin{equation}}
\newcommand{\ee}{\end{equation}}
\newcommand{\bea}{\begin{eqnarray}}
\newcommand{\eea}{\end{eqnarray}}

\def\baselinestretch{1}
\begin{document}
\catcode`@=11
\newtoks\@stequation
\def\subequations{\refstepcounter{equation}%
\edef\@savedequation{\the\c@equation}%
  \@stequation=\expandafter{\theequation}
  \edef\@savedtheequation{\the\@stequation}
  \edef\oldtheequation{\theequation}%
  \setcounter{equation}{0}%
  \def\theequation{\oldtheequation\alph{equation}}}
\def\endsubequations{\setcounter{equation}{\@savedequation}%
  \@stequation=\expandafter{\@savedtheequation}%
  \edef\theequation{\the\@stequation}\global\@ignoretrue

\noindent}
\catcode`@=12
\begin{titlepage}

\title{{\bf  
Scale-independent mixing angles
}}
\vskip2in
\author{  
{\bf J.R. Espinosa$^{1,2}$\footnote{\baselineskip=16pt E-mail: {\tt
espinosa@makoki.iem.csic.es}}} and 
{\bf I. Navarro$^{3}$\footnote{\baselineskip=16pt E-mail: {\tt
ignacio@makoki.iem.csic.es}}}
\hspace{3cm}\\
 $^{1}$~{\small I.M.A.F.F. (CSIC), Serrano 113 bis, 28006 Madrid, Spain}
\hspace{0.3cm}\\
 $^{2}$~{\small I.F.T. C-XVI, U.A.M., 28049 Madrid, Spain}
\hspace{0.3cm}\\
 $^{3}$~{\small I.E.M. (CSIC), Serrano 123, 28006 Madrid, Spain}.
} 
\date{} 
\maketitle 
\def\baselinestretch{1.15} 
\begin{abstract}
\noindent 
A radiatively-corrected mixing angle has to be independent of the choice
of renormalization scale to be a physical observable. At one-loop
in $\ms$, this  occurs for a particular value, $p_*$, of the external 
momentum in the two-point functions used to define the mixing angle: 
$p_*^2=(M_1^2+M_2^2)/2$, where $M_{1,2}$ are the physical masses of
the two mixed particles. We examine
two important applications of this to the Minimal Supersymmetric Standard
Model: the mixing angle for a) neutral Higgs bosons and b) stops. 
We find that this choice of external momentum improves the scale
independence (and therefore provides a more reliable
determination) of these mixing angles.  
\end{abstract}

\thispagestyle{empty}
\vspace{6cm}
\leftline{September 2001}
\leftline{}

\vskip-22cm
\rightline{}
\rightline{IEM-FT-220/01}
\rightline{IFT-UAM/CSIC-01-26}
\rightline{hep-ph/0109126}
\vskip3in

\end{titlepage}
\setcounter{footnote}{0} \setcounter{page}{1}
\newpage
\baselineskip=20pt

\noindent

In Quantum Field Theory renormalized using dimensional regularization and 
minimal subtraction \cite{MS} (or modified minimal subtraction, $\ms$ \cite{MSbar}), 
the parameters of the model at hand (say the Standard Model, SM
or some extension therof) depend on an arbitrary
renormalization scale, $Q$. The running of these parameters with $Q$ is
governed by the corresponding renormalization group equations (RGEs).
Physical observables, on the other hand, cannot depend on the arbitrary scale $Q$,
and the relations that link physical quantities to the running $\ms$ parameters
are of obvious importance in order to make contact between theory and experiment. 

Two familiar examples in the SM concern the Higgs and top quark masses. The relation
between the top-quark pole mass, $M_t$, and the running top Yukawa coupling, $h_t(Q)$,
(or alternatively the running top quark mass) is of the form
\be
h_t(Q)=2^{3/4}G_F^{1/2}M_t\left[1+\delta_t(Q,p^2=M_t^2)
\right]\ ,
\label{mt}
\ee
where $G_F$ is Fermi's constant and the function $\delta_t$, which contains the 
radiative corrections, 
depends explicitly on $Q$ and is evaluated on-shell, {\it i.e.} at external momentum 
satisfying $p^2=M_t^2$. This function $\delta_t$ is known up to 
three loops in QCD \cite{mt,mtQCD} and one loop in electroweak corrections \cite{mtew}.
The relation between the Higgs boson pole mass and its quartic self-coupling $\lambda(Q)$
is similar to (\ref{mt}):
\be
\lambda(Q)={1\over\sqrt{2}}G_F M_h^2\left[1+\delta_h(Q,p^2=M_h^2)
\right]\ .
\label{mh}
\ee
The function $\delta_h$ was obtained at one loop in ref.~\cite{SZ}. 

The scale dependence of the exact functions $\delta_t(Q)$ and $\delta_h(Q)$ in eqs.~(\ref{mt}) and 
(\ref{mh}) must be such that it exactly compensates the scale-dependence of the couplings 
$h_t(Q)$ and $\lambda(Q)$, in such a way that $M_t$ and $M_h$ are scale-independent. 
As shown by these examples, in practice we can only calculate $\delta_t(Q)$ and $\delta_h(Q)$
in some approximation (say up to some finite loop order) and, in general, there is some
residual scale dependence left. In fact, choosing a particular value $Q^*$ of the
renormalization scale by demanding that the residual scale dependence is minimized
(this can be done with different levels of sophistication, see \cite{opti}) 
gives in general a good approximation to the full result, 
or allows a good estimate of higher order corrections. Often there is some physical reason
for the particular value $Q^*$ chosen ({\it e.g.} $Q^*$ might be some average of the
masses of the virtual particles that dominate the loop corrections, or the typical
energy scale of the process studied), but this is not necessarily the case always \cite{mtQCD,fos}.

A physical mass is defined at the pole of the corresponding propagator and therefore
the external momentum in (\ref{mt}) and (\ref{mh}) is set to the physical mass. As we remind
in section~1, this choice ensures that the scale dependence in equations like (\ref{mt}) and 
(\ref{mh}) is the same on both sides, leading to a scale-independent definition of the
physical mass (up to the residual dependences due to higher order corrections just mentioned).
The purpose of this letter is to address the problem of how to obtain
a scale-independent mixing angle between two particles with the same quantum numbers, so that a 
convenient definition of such angles can be achieved. We will show that 
relations similar to (\ref{mt}) and (\ref{mh}) can be found that relate `physical' and 
`running' mixing angles.
Then we show that, at one loop, a scale-independent mixing angle in $\ms$ is possible for a 
very particular choice of external momentum, $p_*$, in the self-energies that contribute
to radiative corrections, with
\be
p_*^2={1\over 2}(M_1^2+M_2^2)\ ,
\ee
where $M_1$ and $M_2$ are the physical masses of the two particles that mix.

In section 1 we present the general derivation of the momentum scale $p_*$ for the
simple case of the mixing angle between two scalar fields. We apply this 
general result to two important cases with phenomenological interest
in the Minimal Supersymmetric Standard Model (MSSM): first to the stop mixing in 
section~2  and then to the mixing between the two ${\cal CP}$-even Higgs scalars
in section~3. We end with some conclusions in section~4.

{\bf 1.} We start by proving the scale-independence (at one-loop) of the pole-mass
for a single scalar field, $\varphi$, with Lagrangian
\bea
{\cal L}&=&{1\over 2}\partial_\mu \varphi_0 \partial^\mu \varphi_0
-{1\over 2}m_0^2\varphi_0^2+...\nonumber\\
&=&{1\over 2}\partial_\mu \varphi \partial^\mu \varphi
-{1\over 2}m^2\varphi^2+{1\over 2}\delta Z_\varphi\partial_\mu \varphi \partial^\mu
\varphi 
-{1\over 2}\delta m^2 \varphi^2+...\ ,
\eea
where $\varphi_0,m_0$ ($\varphi,m$) are bare (renormalized, say in $\ms$-scheme) 
quantities and $\delta Z_\varphi,\delta m^2$ are counterterms. They are related by
\bea
\varphi_0&=&\left(1+{1\over 2}\delta Z_\varphi\right)\varphi\ ,\\
m_0^2&=&m^2+\delta m^2-\delta Z_\varphi m^2\ .
\eea
The quantities $\varphi$ and $m^2$ depend on the renormalization scale $Q$
through $\delta Z_\varphi$ and $\delta m^2$:
\bea
{d\ln\varphi\over d\ln Q^2}&\equiv & \gamma =
-{1\over 2}{d \delta Z_\varphi\over d\ln Q^2}\ ,\\
{d m^2\over d\ln Q^2}&\equiv & \beta_{m^2}=
-{d \delta m^2\over d\ln Q^2}+m^2{d \delta Z_\varphi\over d\ln Q^2} 
\equiv \beta_{m^2}^0-2\gamma m^2 \label{betam}\ .
\eea
The relation between the one-loop bare and renormalized inverse propagators,
$\Gamma_0(p^2)$ and $\Gamma(p^2)$ respectively, is
\be
\Gamma(p^2)\equiv p^2 - m^2 + \Pi(p^2)= (1+\delta Z_\varphi) \Gamma_0(p^2)=
(1+\delta Z_\varphi) [p^2 - m_0^2 + \Pi_0(p^2)]\ ,
\label{invpr}
\ee
where $\Pi_0(p^2)$ $[\Pi(p^2)]$ is the bare (renormalized) one-loop self-energy
for external momentum $p^2$. From (\ref{invpr}) we can obtain the renormalization-scale
dependence of the $\ms$-renormalized self-energy:
\be
{\partial \Pi(p^2)\over \partial\ln Q^2}= \beta_{m^2}^0-2\gamma p^2\ .
\label{betapi}
\ee
The physical mass, $M$, is defined as the real part of the propagator 
pole\footnote{Throughout the paper we will not be interested in particle decay widths
and we ignore the imaginary part of the self-energies involved.}. Therefore, $M$ is
given by
\be
M^2\equiv m^2-\Pi(p^2=M^2)\ .
\ee
{}From this, using (\ref{betam}) and (\ref{betapi}) we find
\be
{d M^2\over d\ln Q^2}={d m^2\over d\ln Q^2}-{d \Pi(M^2)\over d\ln Q^2}=
\left.2\gamma(m^2-p^2)\right|_{p^2=M^2}+{\cal{O}}(\hbar^2)\ .
\label{dMdQ}
\ee
For the on-shell choice $p^2=M^2$, noting that $M^2=m^2+{\cal{O}}(\hbar)$,
eq.~(\ref{dMdQ}) is zero at one-loop order (of course the proof can be 
extended to all orders).

The scale-independence just proved also holds in the case of mixed fields.
Consider two scalar fields, $\varphi_1$ and $\varphi_2$, with the
same quantum numbers, so that they can mix. Their inverse propagator, for
external momentum $p$, is a $2\times 2$ matrix which at one-loop order has
the form
\begin{equation}
{\bf \Gamma}(p^2)\
\equiv
p^2\ {\bf I_2}-[{\bf m^2}-{\bf \Delta}(p^2)]=
\left[\begin{array}{cc}
p^2 - m_{11}^2 + \Delta_{11} & -m_{12}^2 + \Delta_{12}\\[2mm]
-m_{21}^2 + \Delta_{21} & p^2 - m_{22}^2 + \Delta_{22}
\end{array}\right]\ .
\label{invp}
\end{equation}  
where ${\bf m^2}$ is the $\ms$-renormalized (`tree-level') 
mass matrix and ${\bf \Delta}(p^2)$ contains the
one-loop radiative corrections. In $\ms$-scheme (or 
for supersymmetric theories $\dr$ \cite{DRbar}, the SUSY version of $\ms$, with dimensional
reduction instead of dimensional regularization), the elements of ${\bf m^2}$ depend implicitly 
on the renormalization scale $Q$ through an equation of the form:
\be
{d m_{ij}^2\over d\ln Q^2}\equiv\beta_{ij}\equiv
\beta_{ij}^0 - \sum_k\left(\gamma_{ik}
m_{kj}^2+\gamma_{kj} m_{ik}^2\right) .
\label{RGEm}
\ee
In (\ref{RGEm}) we have written separately the contributions from wave-function
renormalization, with the anomalous dimensions $\gamma_{ij}$ defined by
\be
{d\varphi_i\over d\ln Q^2}\equiv \sum_k\gamma_{ik} \varphi_k\ . 
\ee
The fact that $\gamma_{ij}$ is in general a $2\times 2$ matrix reflects the possibility
of having kinetic mixing between the two scalar fields. 

The one-loop radiative corrections to the inverse propagator, collected in 
${\bf \Delta}$, depend explicitly on the external momentum $p$ and on 
$\ln Q^2$. In fact, the elements of that matrix satisfy 
\be
{\partial \Delta_{ij}\over\partial \ln
Q^2}=\beta_{ij}^0- 2 p^2\gamma_{ij}\ .
\label{RGEDm}
\ee

The two mass eigenvalues are the poles of ${\bf \Gamma}^{-1}(p^2)$,
that is, the solutions $p^2=M^2_i$ ($i=1,2$) of the equation 
\be
{\mathrm Det}\ [Re\ {\bf \Gamma}(p^2)]=p^4-p^2\ {\mathrm Tr}\ [{\bf
M^2}(p^2)] +{\mathrm Det}\ [{\bf M^2}(p^2)]=0\ ,
\label{pole}
\ee
where ${\bf M^2}(p^2)\equiv{\bf m^2}-{\bf \Delta}(p^2)$. To find $M_{1,2}^2$, 
eq.~(\ref{pole}) has to be solved self-consistently to the order we work 
(one-loop). It is easy to show that the mass eigenvalues, $M^2_i$, 
are indeed scale-independent at one loop. Taking the derivative of 
eq.~(\ref{pole}) with respect to $\ln Q^2$, with $p^2=M_i^2$, we find 
\be
{d M_i^2\over d\ln Q^2}\propto \left[-M_i^2 {d\ {\mathrm Tr}\ {\bf M^2}\over
d\ln
Q^2}+ {d\ {\mathrm Det}\ {\bf M^2}\over d\ln Q^2}\right]\ .
\label{proof}
\ee
Using eqs.~(\ref{RGEm}) and (\ref{RGEDm}) to evaluate $d M_{ij}^2/d\ln Q^2$
and making a loop expansion, the right-hand side of eq.~(\ref{proof}) is shown to be 
proportional to
\be
m_i^4-m_i^2\  {\mathrm Tr} \ {\bf m^2}
+{\mathrm Det}\ {\bf m^2}\ ,
\label{zero}
\ee
where $m_i^2$ is an eigenvalue of the tree-level mass matrix, and therefore
must satisfy the secular equation ${\mathrm Det}[m_i^2{\bf I_2}-{\bf m^2}]=0$, 
so that (\ref{zero}) is zero [and then also (\ref{proof}) vanishes]. 
This establishes the scale independence
of $M_i^2$ at one-loop. In summary, this agrees nicely with the well known result
that a physical definition of the radiatively
corrected masses, $M_i^2$, requires the self-energy corrections 
$[{\bf\Delta}(p^2)]$ to be evaluated on-shell, {\it i.e.} at $p^2=M_i^2$.

Besides the particle masses, ${\bf \Gamma}(p^2)$ in eq.~(\ref{invp}) contains 
also information on the mixing between the two particles.
We can define the radiatively-corrected mixing angle, $\alpha(p^2)$, as the
angle of the rotation that diagonalizes the mass matrix ${\bf M^2}=
{\bf m^2}-{\bf\Delta}(p^2)$:
\be
\tan [2\alpha(p^2)] \equiv {\sqrt{4
\left[m_{12}^2-\Delta_{12}(p^2)\right]
\left[m_{21}^2-\Delta_{21}(p^2)\right]}\over
m_{11}^2-m_{22}^2+\Delta_{22}(p^2)-\Delta_{11}(p^2)}\ .
\label{mixang}
\ee
Proceeding like we did for the masses, the scale-dependence of 
$\tan [2\alpha(p^2)]$ can be extracted from
eqs.~(\ref{RGEm},\ref{RGEDm}). At one-loop order, it is simply given by
\be
{d \tan 2\alpha\over d\ln Q^2}\propto \left[2p^2- {\mathrm
Tr} \ {\bf m^2}\right]\ .
\label{proofp}
\ee
From this result we conclude that a scale-independent mixing angle 
can be defined at external momentum 
\be
p_*^2\equiv{1\over 2}(M_1^2+M_2^2)\ .
\label{pstar}
\ee
This is the simple result we wanted to prove. We choose to define $p_*$ in terms 
of the radiatively corrected masses in analogy with the on-shell definition
of physical masses. Eq.~(\ref{proofp}) involves 
tree-level masses and
cannot be used to justify this choice, although it is of course compatible 
with it at one-loop. Besides the analogy with OS masses, numerical examples 
in later sections will provide good support for this choice.

In applying the previous prescription to gauge theories one should 
worry about the gauge-independence of the mixing-angle definition.
We expect our prescription to be amenable of improvement in order to
make it also gauge-independent (along the lines of \cite{yamada}). The
results of such analysis will be presented elsewhere \cite{eny}. For
our current purposes notice that in the examples of the 
following sections the scale dependence of the parameters that 
enter the definition of the mixing angle is very mildly affected by electroweak
gauge couplings, so that it is a good approximation for our numerical analyses to 
neglect them.

{\bf 2.} In the context of the MSSM, one particular case in which the
previous discussion is of interest concerns the stop sector. This sector 
consist of two scalars, $\tilt_{L,R}$, supersymmetric partners of the top quark 
which, after electroweak symmetry breaking, can mix. Their tree-level mass matrix 
is given by
\begin{equation}
{\bf m}^2_{\tilt}\
 \simeq\
\left[\begin{array}{cc}
M_L^2+m_t^2  &  m_t X_t\\[2mm]
m_t X_t^*      &  M_R^2+m_t^2
\end{array}\right]\ ,
\label{stopmat}
\end{equation}  
where $M_L$ ($M_R$) is the squared soft mass for $\tilt_{L}$
($\tilt_{R}$), $m_t$ is the top mass and $X_t\equiv A_t+\mu^*/\tan\beta$,
with $A_t$ the soft trilinear coupling associated to the top Yukawa coupling,
$\mu$ the Higgs mass parameter in the superpotential and $\tan\beta$ the
ratio between the vacuum expectation values of the two Higgs doublets of the model.
In (\ref{stopmat}) we have neglected gauge couplings, which give a small
contribution through $D$-terms.  The one-loop self-energy corrections 
for stops in the MSSM, ${\bf \Delta}_{\tilt}(p^2)$, can be found in \cite{PBMZ,Donini}.

The stop mixing angle, $\theta_t(p^2)$, including one-loop radiative corrections, can be
defined for any value of the external momentum $p$ as the angle of the basis rotation 
which diagonalizes ${\bf M}^2_{\tilt}\equiv {\bf m}^2_{\tilt}-{\bf \Delta}_{\tilt}(p^2)$. 
Following section~1, we define a scale-independent mixing angle 
${\tilde \theta}_t\equiv\theta_{t}(p_*^2)$, for 
$p_*^2=(M_{\tilt_1}^2+M_{\tilt_2}^2)/2$, where $M_{\tilt_i}^2$ are the (one-loop) stop
mass eigenvalues. 

In figures 1.a and 1.b, we illustrate the scale dependence of $\tan
2\theta_t(p^2)$ in the range 100 GeV $\leq Q \leq$ 1 TeV 
for different values of the external momentum $p$. We choose
the following stop parameters: $M_L= 1500$ GeV, $M_R=300$ GeV, $X_t=1$ TeV and
$\tan\beta=10$ for figure 1.a, and $M_L=M_R=X_t= 1$ TeV,
$\tan\beta=10$ for figure 1.b (these are values at the electroweak scale, 
$Q\sim 100$ GeV).  
We work in the approximation of neglecting all couplings other than the
strong gauge coupling and the top and bottom Yukawa couplings for the RG
evolution of the tree-level matrix (\ref{stopmat}). The scale evolution of
this matrix is considered only at one-loop leading-log order, that is
\be
m_{ij}^2(Q)\simeq m_{ij}^2(Q_0)+\beta_{ij}\ln{Q^2\over Q_0^2}\ ,
\label{LL}
\ee
with $Q_0=100$ GeV. The quantities $\beta_{ij}$ depend also on the soft mass
of right-handed sbottoms, $M_D$, on the soft trilinear coupling, $A_b$, 
associated to the bottom Yukawa coupling in the superpotential and on the
gluino mass, $M_g$. 
We take $M_D=M_g=1$ TeV and $A_b=0$.
The upper (lower) curves correspond to $p^2$ equal to the heavier (lighter) mass
eigenvalue (of the radiatively-corrected mass matrix) while the flat curves 
in-between have
$p^2=p_*^2\equiv(M_{\tilt_1}^2+M_{\tilt_2}^2)/2$. The improvement in scale
independence is dramatic when this last choice is made. The difference in
$\tan 2\theta_t$ due to different choices of the external momentum can be 
a $\sim 5-7\%$ effect in the case of figure 1.a and much larger for 1.b
(depending strongly on the value chosen for the renormalization scale).
\begin{figure}
\vspace{1.cm}
\centerline{\hbox{
\psfig{figure=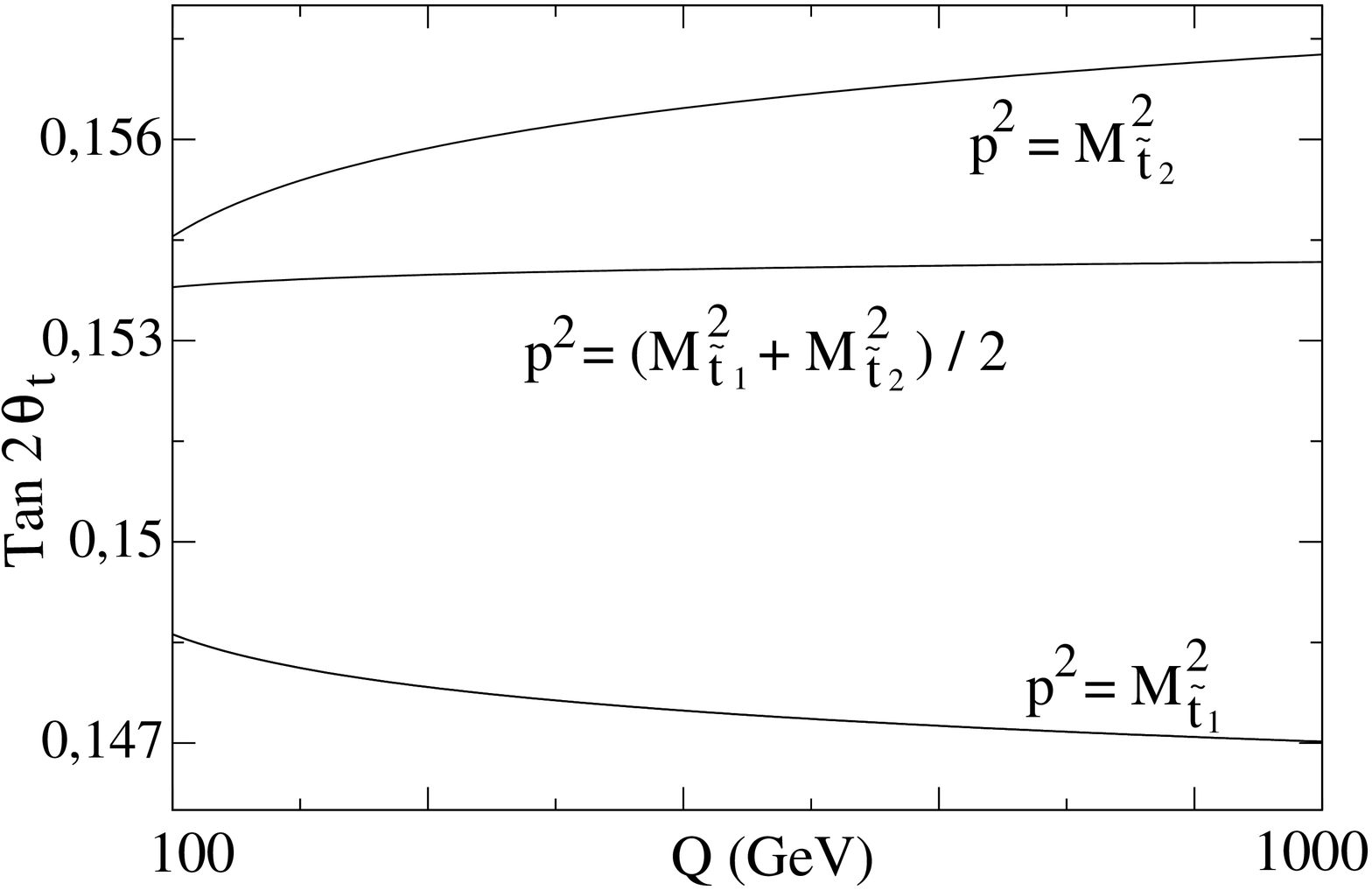,height=6cm,width=6cm,angle=-90,bbllx=4.cm,%
bblly=6.cm,bburx=20.cm,bbury=28.cm}
\psfig{figure=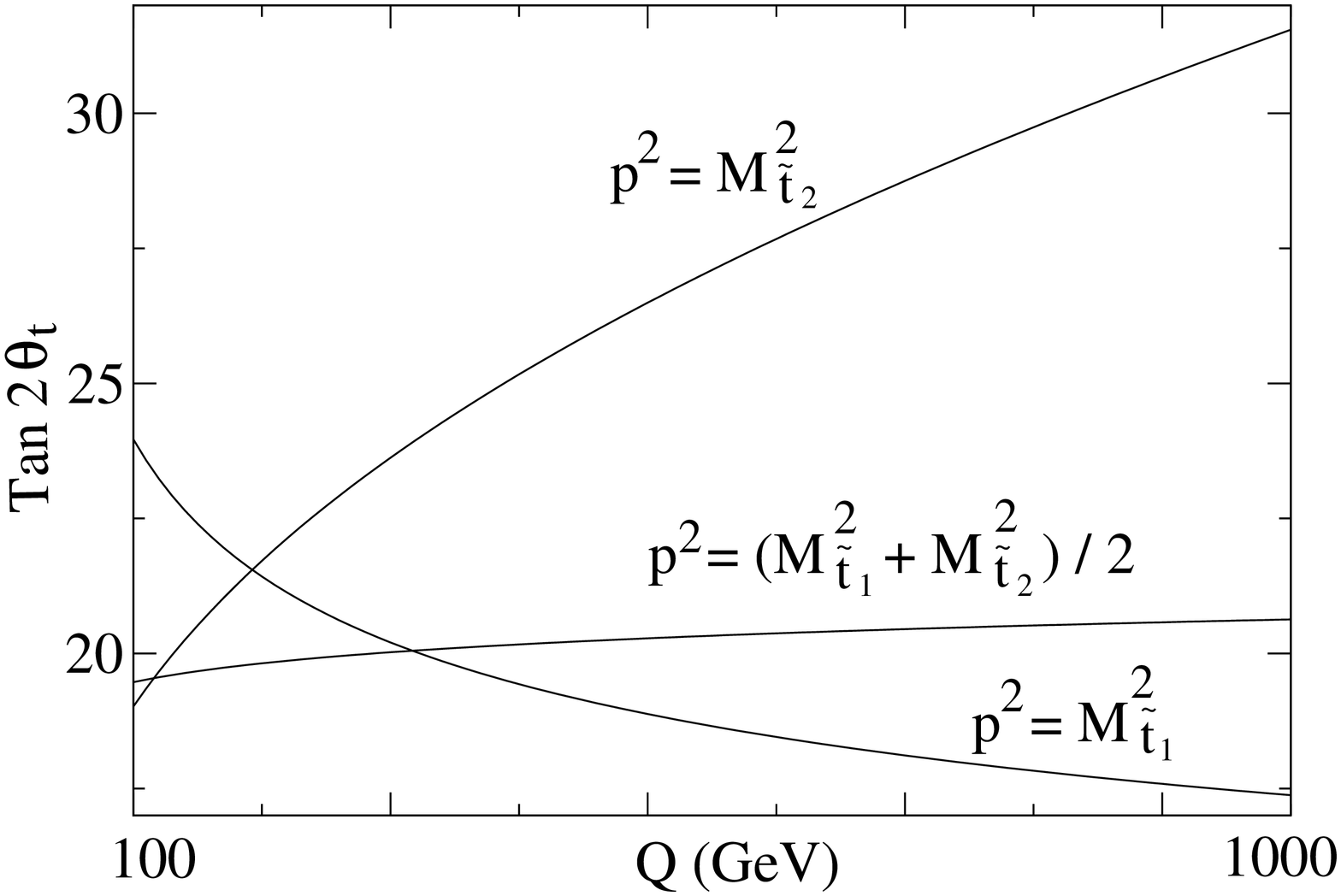,height=6cm,width=6cm,angle=-90,bbllx=4.cm,%
bblly=0.cm,bburx=20.cm,bbury=22.cm}}}
\caption
\noindent{\footnotesize 
Scale dependence of the stop mixing-angle $\theta_t(p^2)$
for different values of the external momentum $p$ as indicated.
Figure 1.a (left
plot) has $M_L= 1500$ GeV, $M_R=300$ GeV, $Xt=1$ TeV 
while figure 1.b (right plot) has $M_L=M_R=X_t= 1$ TeV. In both
cases $\tan\beta=10$.}
\end{figure}

Once a scale-independent mixing angle ${\tilde \theta}_t$  has been
obtained, we can also define a scale-independent, 'on-shell',
stop mixing parameter, $X_t^{OS}$, by the relation (already used in \cite{en})
\be
\sin 2 {\tilde \theta}_t\equiv {2M_tX_t^{OS} \over M_{\sto}^2-M_{\stw}^2}\ ,
\ee
where $M_t$ is the pole mass of the top quark.

{\bf 3.} Another sector of the MSSM in which radiative corrections and 
mixing effects are quite
relevant is the Higgs sector, in particular for the mixing
between the two neutral Higgses\footnote{For an incomplete list of
previous literature on  the radiatively corrected
Higgs mixing angle see {\it e.g.} refs.~\cite{alphaos} in on-shell scheme
and refs.~\cite{PBMZ,andrea,alphadr} in $\dr$-scheme.} 
(we assume ${\cal CP}$ is approximately conserved in
the Higgs sector so that only two ${\cal CP}$-even states can mix. If this were
not the case, similar analyses would be possible). In the $\{H_1^0, H_2^0\}$ basis,
the Higgs mass matrix, corrected at one loop, is
\begin{equation}
{\bf M}^2_{H}\
 \simeq\
\left[\begin{array}{cc}
m_A^2 s_\beta^2+m_Z^2 c_\beta^2-\Delta_{11}  &-(m_A^2+m_Z^2)s_\beta
c_\beta -\Delta_{12} \\[2mm]
-(m_A^2+m_Z^2)s_\beta c_\beta -\Delta_{21}   & m_A^2 c_\beta^2+m_Z^2
s_\beta^2-\Delta_{22}   
\end{array}\right]\ .
\label{higgsmat}
\end{equation}  
In this expression, $m_Z$ and $m_A$ are the $\dr$ masses of the $Z^0$ gauge
boson and the pseudoscalar Higgs, $A^0$, respectively and we use the 
shorthand notation $s_\beta=\sin\beta$, etc. 
The diagonal one-loop self-energy corrections
include a piece from Higgs tadpoles (see {\it e.g.} ref.~\cite{PBMZ})
which ensure that Higgs vacuum expectations values minimize the one-loop
effective potential. With this definition, the parameter $\tan\beta$
has the usual RGE in terms of the Higgs anomalous dimensions\footnote{
Other definitions of $\tan\beta$ are possible, see {\it e.g.} 
\cite{Yamada}.}:
\be
{d\tan\beta\over d\ln Q^2}\equiv \gamma_2-\gamma_1\ ,
\ee
with $\gamma_i\equiv d\ln H_i^0/d\ln Q^2$. In spite of
this small complication present in the Higgs sector, the general derivation
given in section~1 gets through also in this case: we get a 
scale-independent Higgs mixing angle, $\alpha$, with the external momentum  
taken as $p^2=p_*^2=(M_h^2+M_H^2)/2$, where $M_h$ and $M_H$ are the masses 
of the light and heavy ${\cal CP}$-even Higgses respectively\footnote{The 
momentum dependence of the Higgs mixing angle was also addressed in \cite{marco}.}.

To show this, we plot $\tan [2\alpha(p^2)]$ in figures 2 and 3 as a function of the
renormalization scale, with 100 GeV $\leq Q\leq$ 1 TeV, for different choices of 
parameters. In figure~2 we take $m_A=500$ GeV, with $\tan\beta=3$ in the 
upper-left plot, $\tan\beta=10$ in the upper-right one and $\tan\beta=40$
in the lower plot. 
For the rest of parameters we take $X_t=0$, $X_b\equiv A_b+\mu\tan\beta=0$
({\it i.e.} diagonal squark masses), $Y_t\equiv A_t-\mu\tan\beta=0$,
and $M_L=M_R=M_D=1$ TeV.
Just like in the previous section, we have made several approximations for
our numerical examples: we neglect all couplings other than the top and
bottom Yukawa couplings (here the strong coupling does not enter in 
one-loop corrections) and consider the RG evolution of the tree-level
piece of (\ref{higgsmat}) only at one-loop leading-log level [like in
eq.~(\ref{LL})].

\begin{figure}[t]
\vspace{1.6cm}
\centerline{\hbox{
\psfig{figure=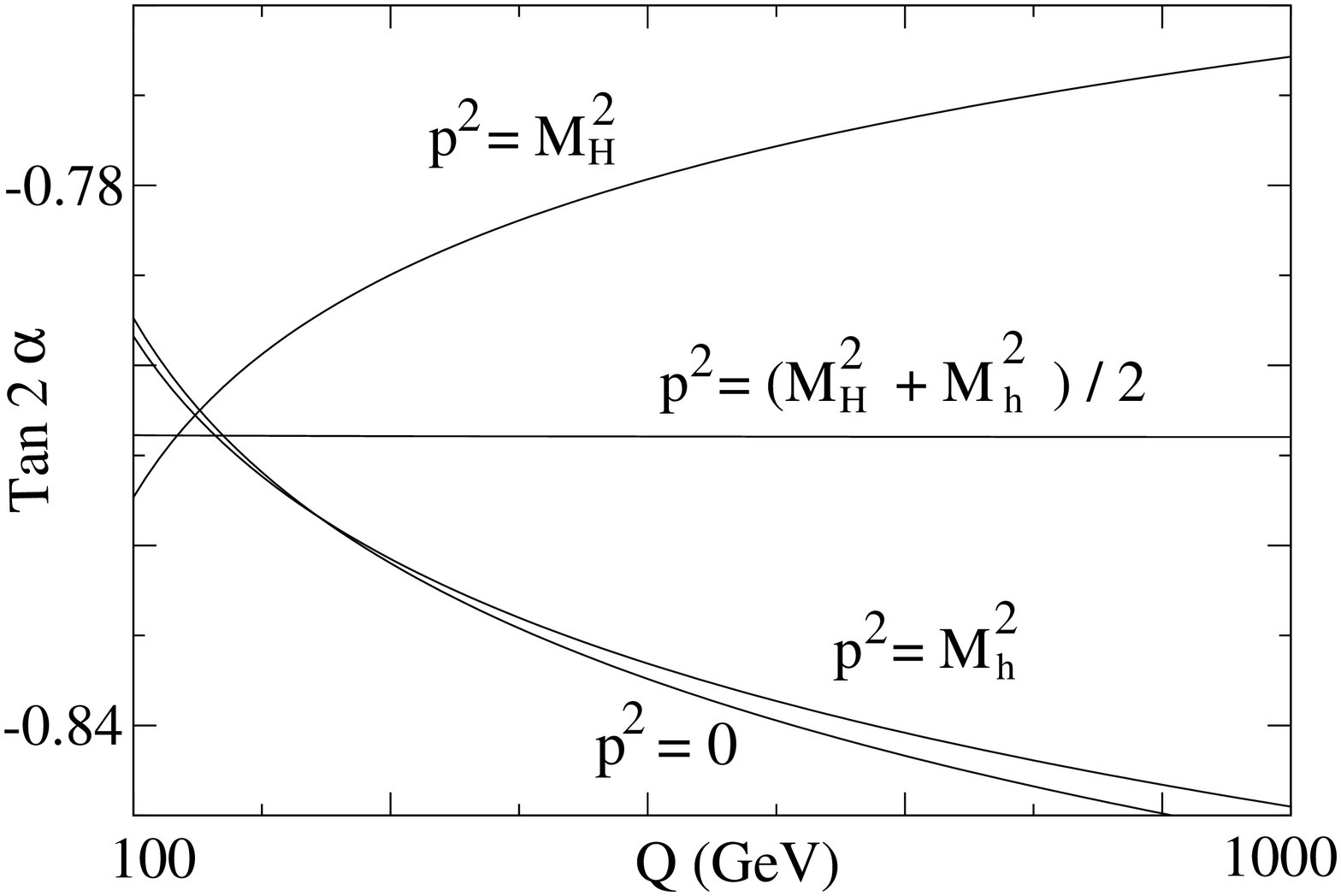,height=6.cm,width=6cm,angle=-90,bbllx=7.cm,%
bblly=6.cm,bburx=24.cm,bbury=28.cm}
\psfig{figure=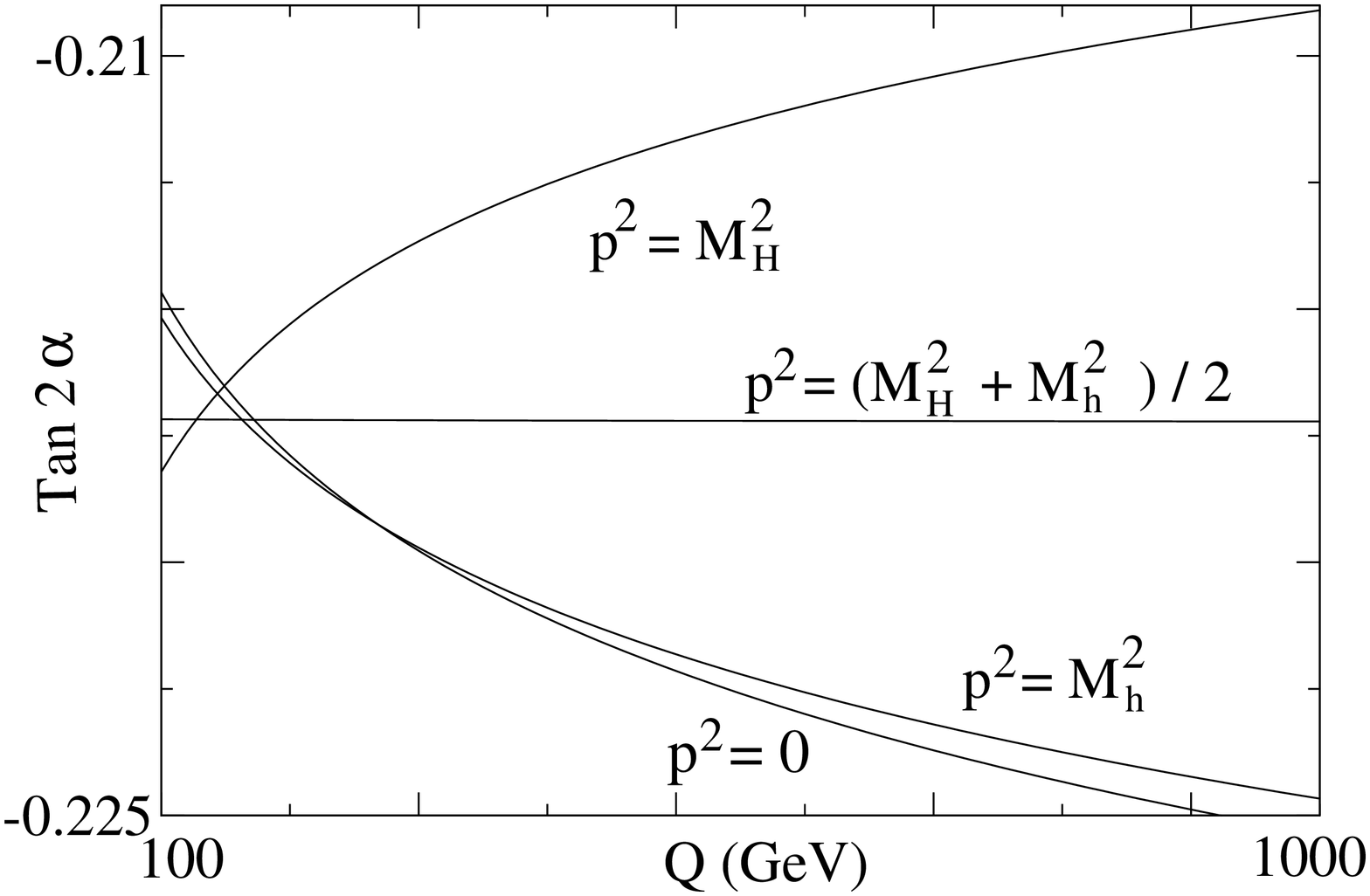,height=6.cm,width=6cm,angle=-90,bbllx=7.cm,%
bblly=0.cm,bburx=24.cm,bbury=22.cm}}}
\centerline{\hbox{
\psfig{figure=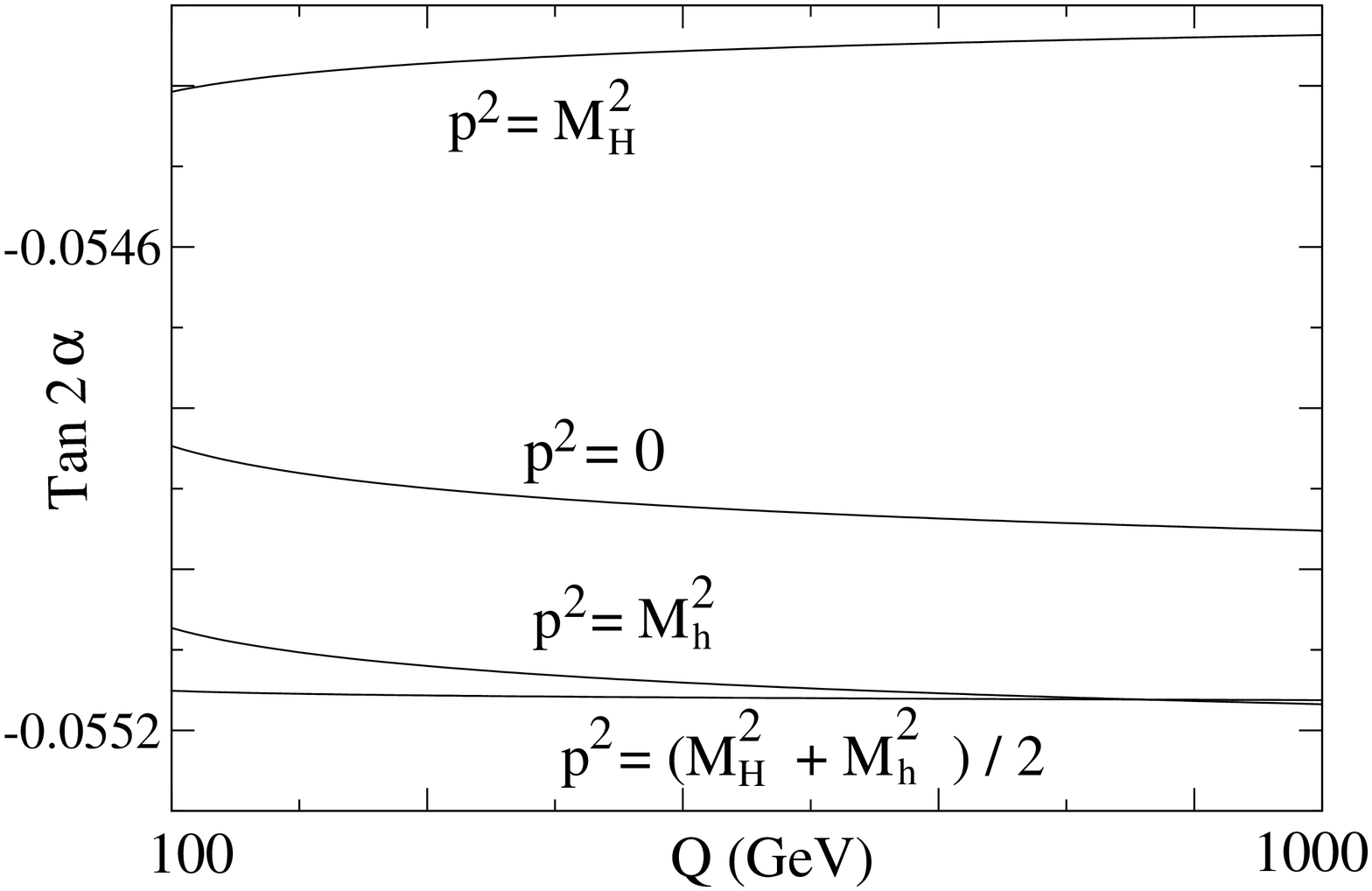,height=6.cm,width=6cm,angle=-90,bbllx=3.5cm,%
bblly=4.cm,bburx=20.5cm,bbury=26.cm}}}
\caption
 \noindent{\footnotesize 
Scale-dependence of the momentum-dependent Higgs mixing-angle $\alpha(p^2)$
for $p^2$ values as indicated. All figures have $m_A=500$ GeV,  $X_t=X_b=Y_t
=0$ and $M_L=M_R=M_D=1$ TeV, while $\tan\beta$ is 3 for the upper-left plot, 
10 for the upper-right one and 40 for the lower plot.}
\end{figure}

{}From figure~2 we see that the choice $p^2=p_*^2=(M_h^2+M_H^2)/2$ indeed improves 
significantly the scale-independence of the Higgs mixing angle $\alpha(p^2)$.
In these plots we compare this choice of external momentum with three other 
possibilities: $p^2$ equal to one of the Higgs masses squared ($M_h^2$ or $M_H^2$,
radiatively corrected) or $p^2=0$. The possibility $p^2=M_h^2$ 
has been used frequently in the
literature to define the Higgs mixing angle. The same can be said of
$p^2=0$, which is the choice made when radiative corrections are extracted
from the effective potential only, without correcting for wave-function
renormalization effects. 
\begin{figure}
\vspace{2cm}
\centerline{\hbox{
\psfig{figure=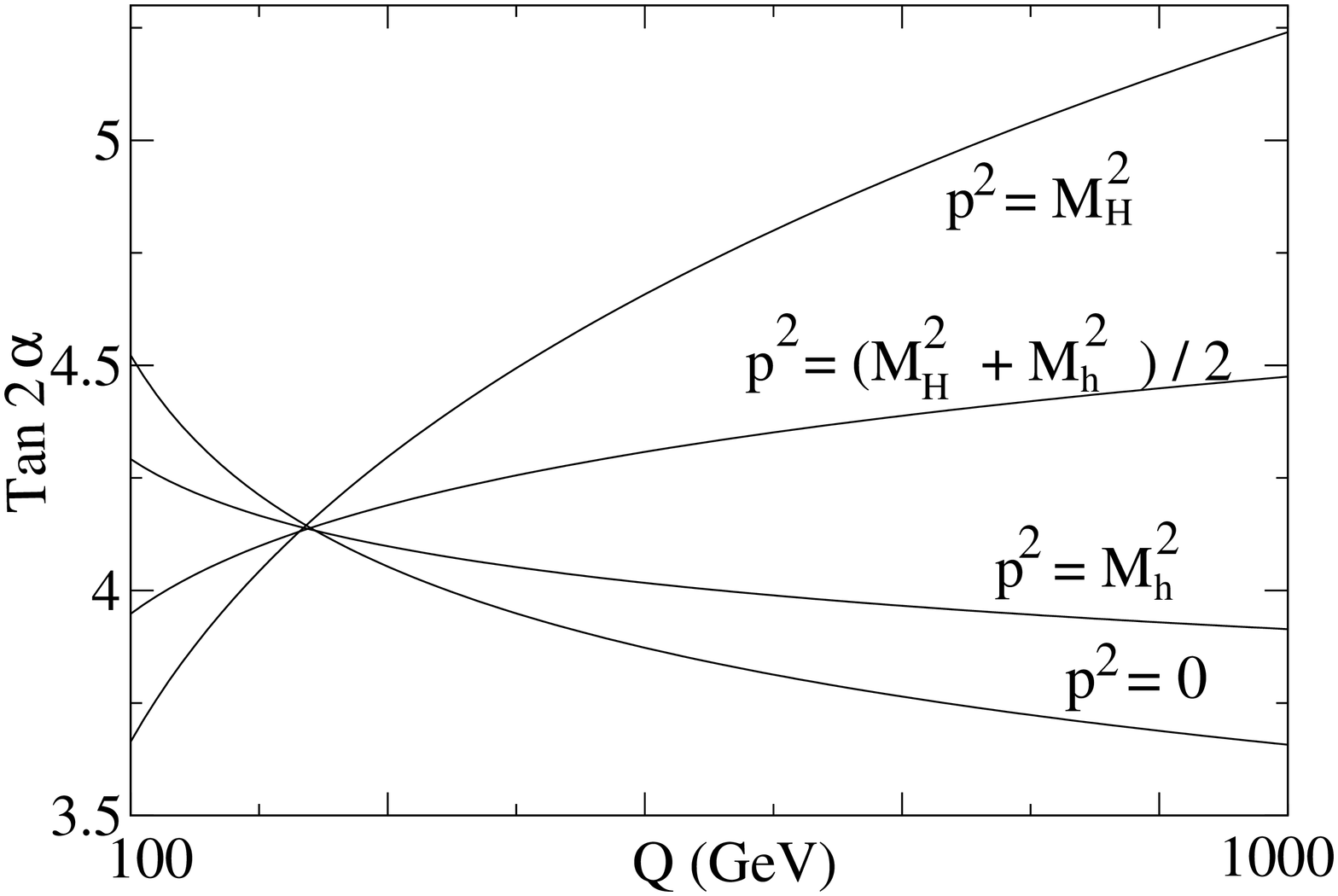,height=6.cm,width=6cm,angle=-90,bbllx=8.cm,%
bblly=6.cm,bburx=25.cm,bbury=28.cm}
\psfig{figure=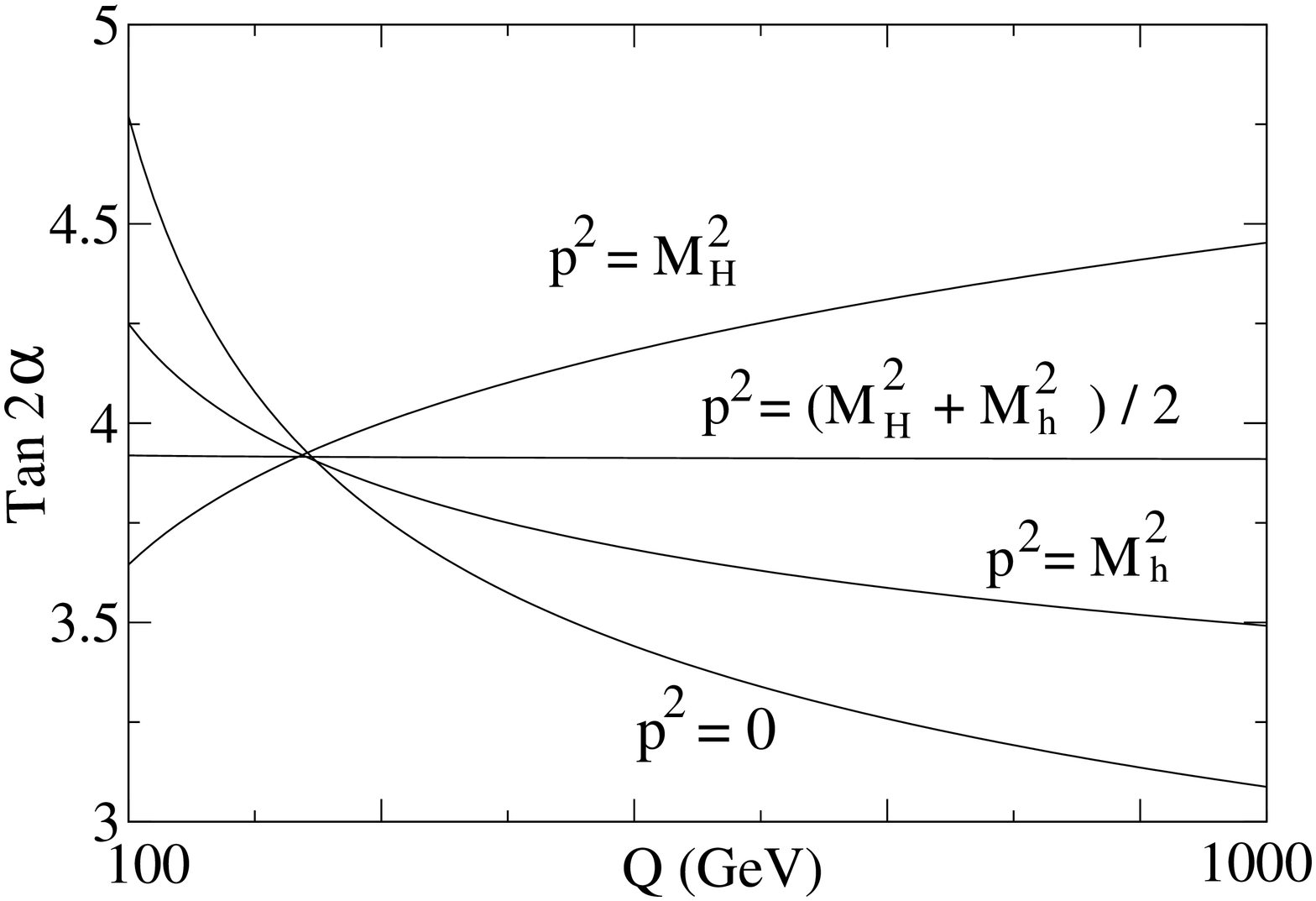,height=6.cm,width=6cm,angle=-90,bbllx=8.cm,%
bblly=0.cm,bburx=25.cm,bbury=22.cm}}}
\centerline{\hbox{
\psfig{figure=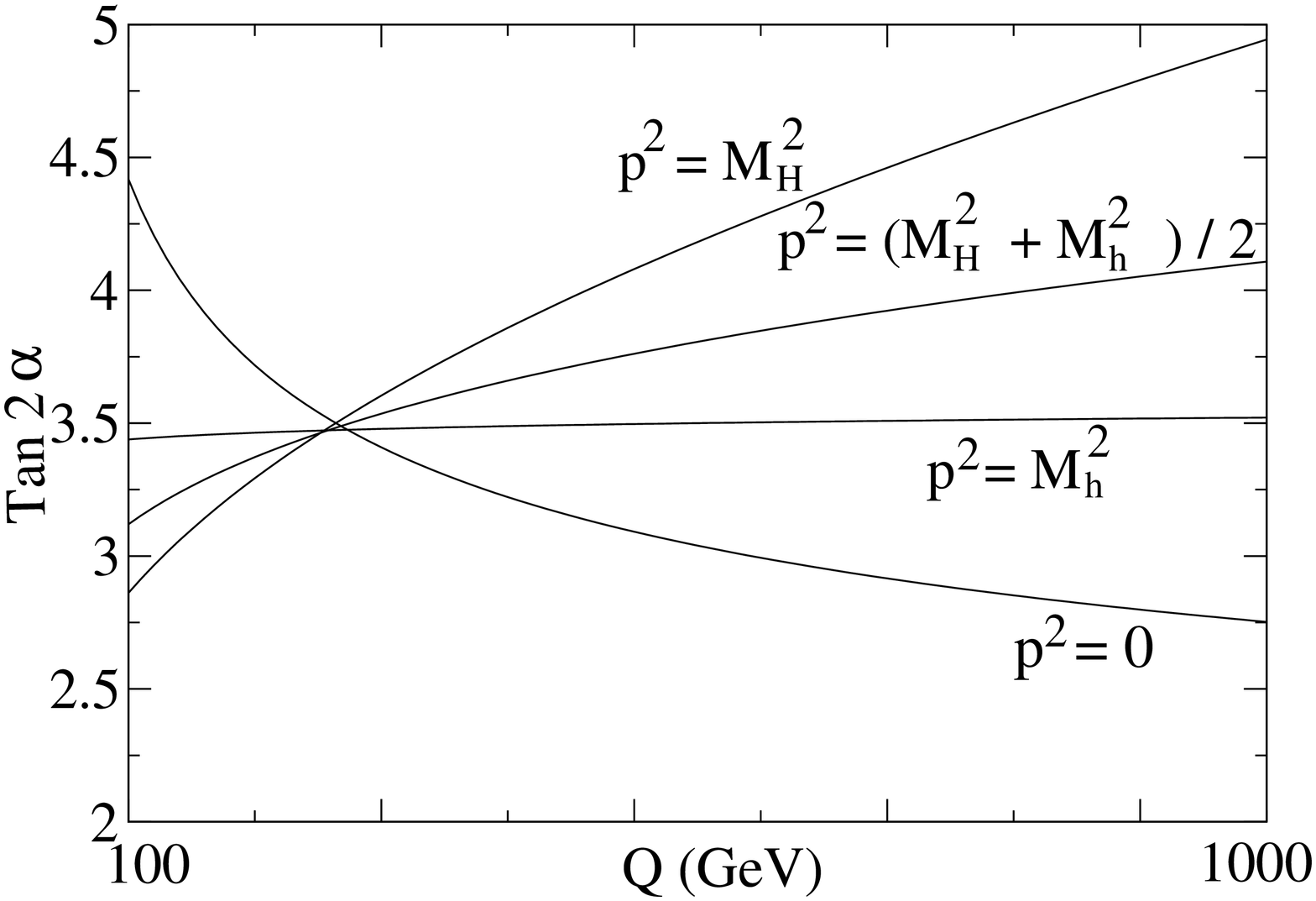,height=6.cm,width=6cm,angle=-90,bbllx=4.5cm,%
bblly=6.cm,bburx=21.5cm,bbury=28.cm}
\psfig{figure=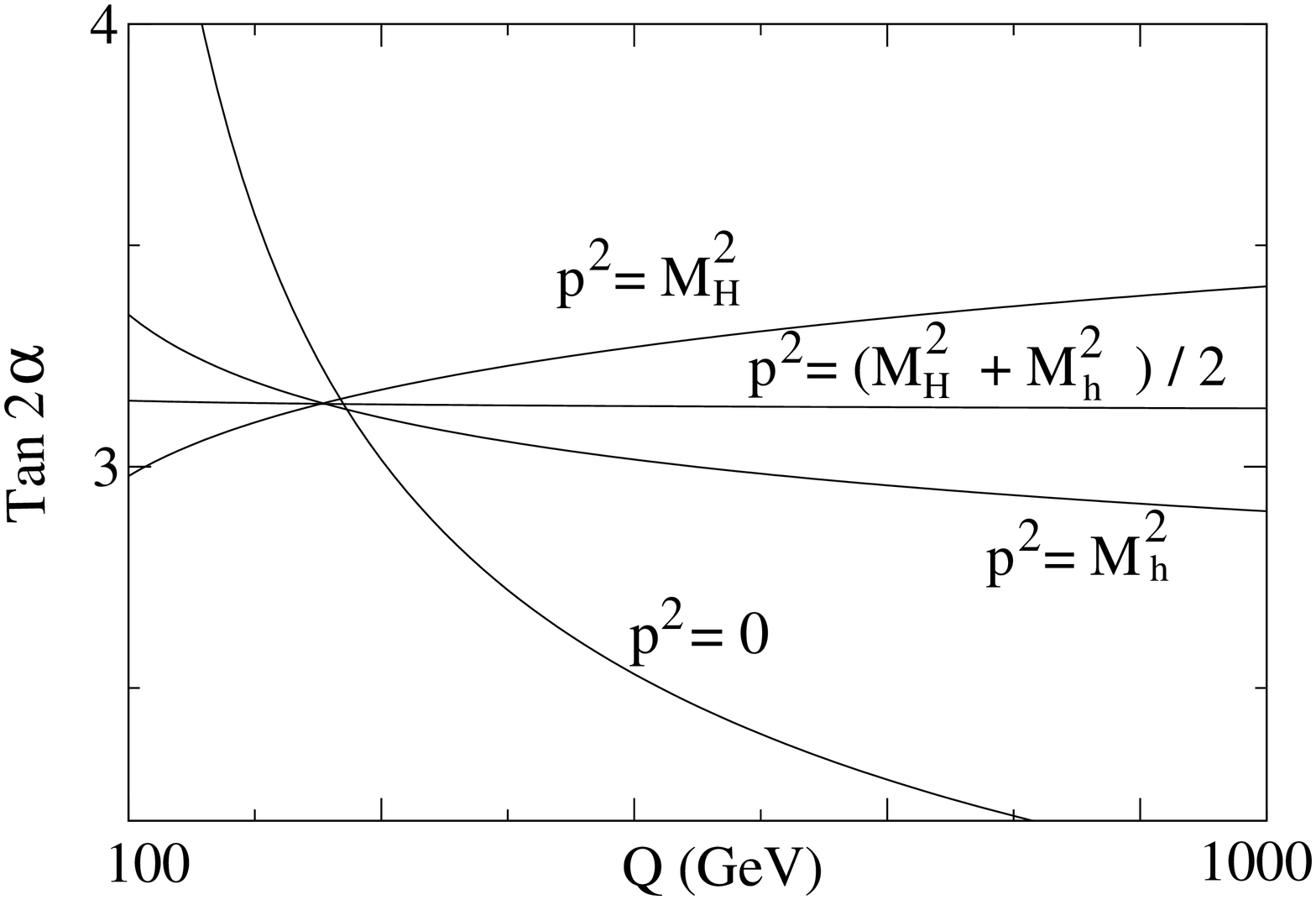,height=6.cm,width=6cm,angle=-90,bbllx=4.5cm,%
bblly=0.cm,bburx=21.5cm,bbury=22.cm}}}
\centerline{\hbox{
\psfig{figure=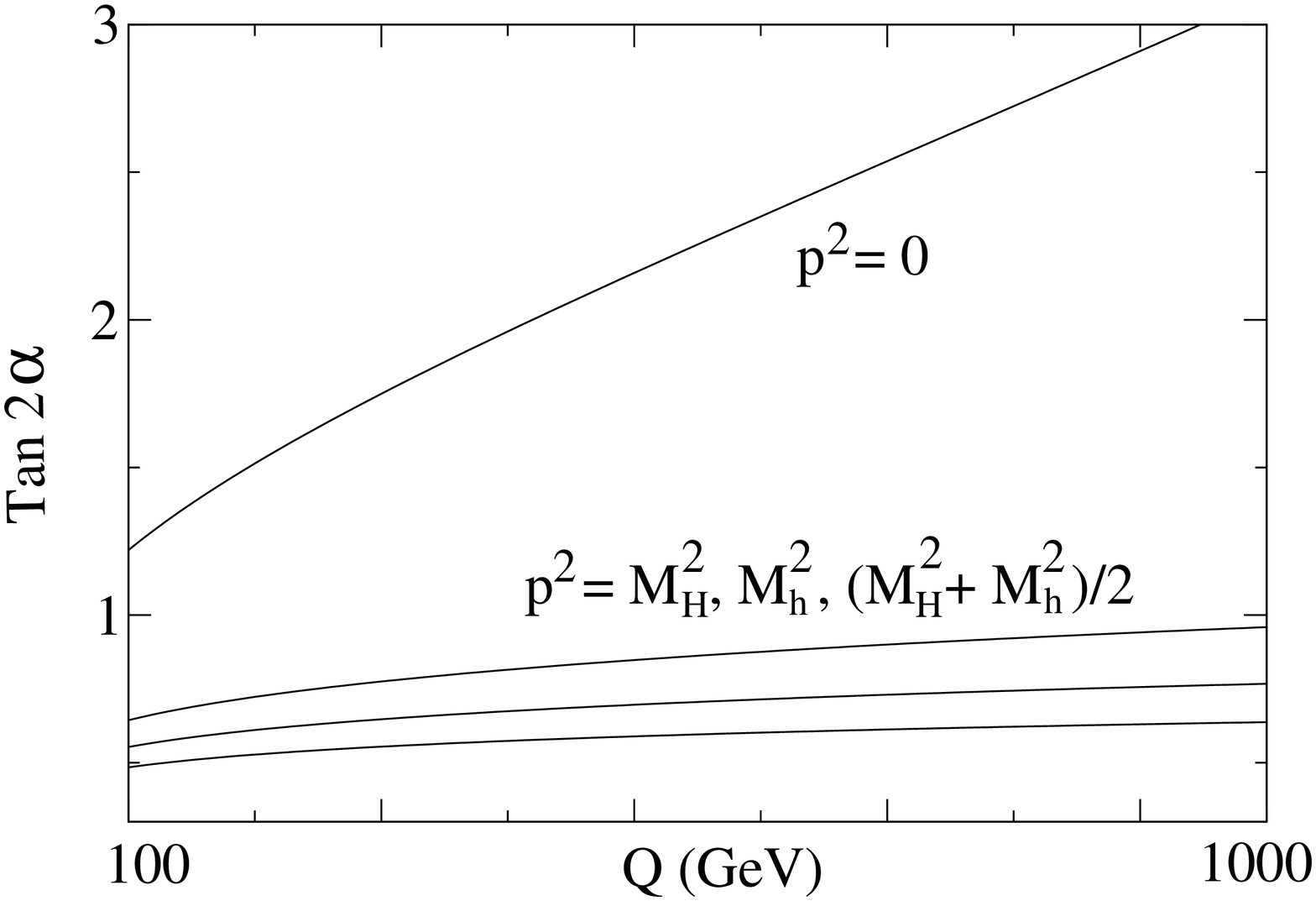,height=6.cm,width=6cm,angle=-90,bbllx=1.5cm,%
bblly=6.cm,bburx=18.5cm,bbury=28.cm}
\psfig{figure=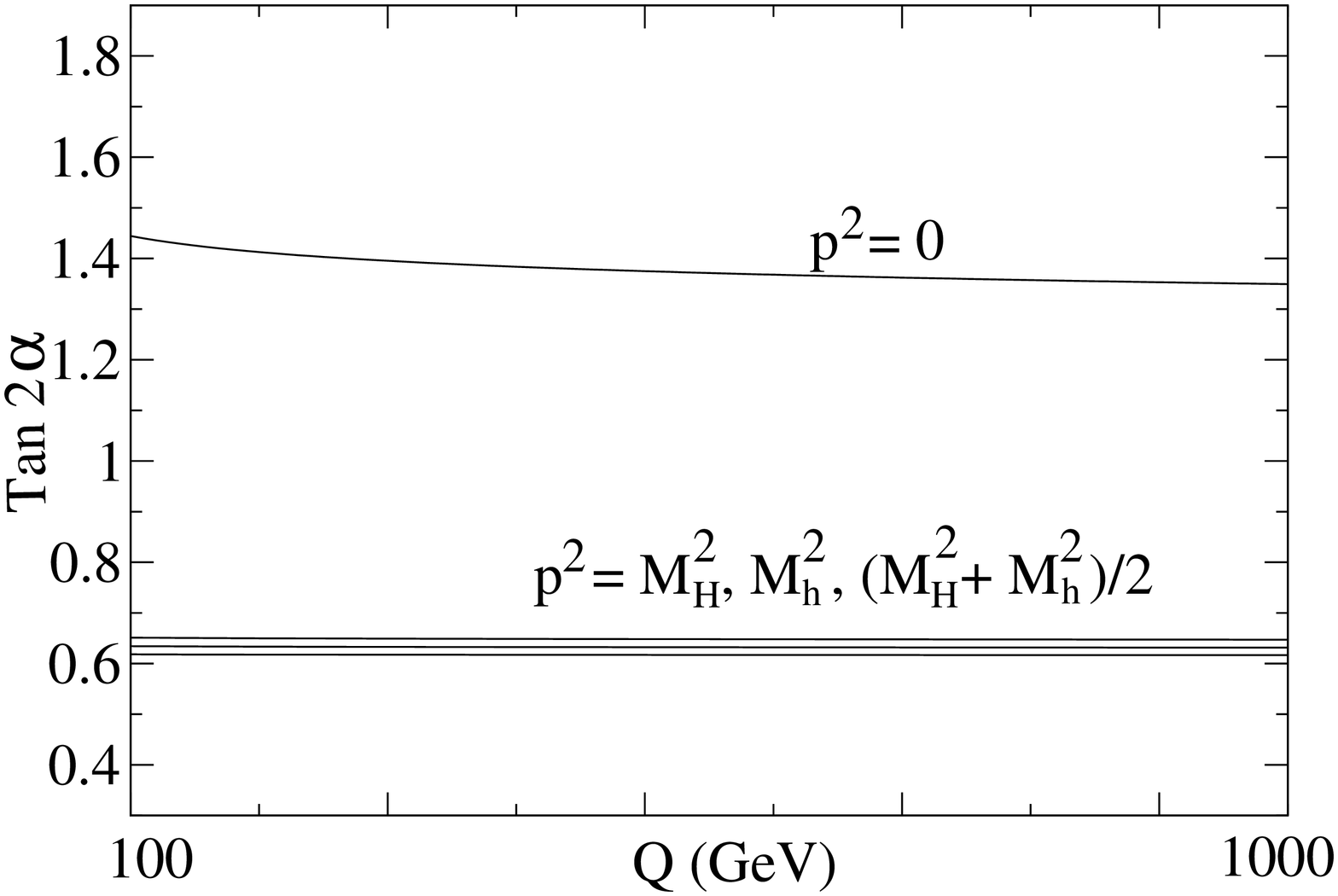,height=6.cm,width=6cm,angle=-90,bbllx=1.5cm,%
bblly=0.cm,bburx=18.5cm,bbury=22.cm}}}
\caption
 \noindent{\footnotesize 
Scale dependence of the momentum-dependent Higgs mixing-angle $\alpha(p^2)$
for $p^2$ values as indicated. All figures have $m_A=120$ GeV;
the parameter $\tan\beta$ is $3,10,40$ from top to bottom plots
while $X_t=X_b=Y_t=0$ and $M_L=M_R=M_D=1$ TeV. Left plots use a purely one-loop
definition of $\alpha$ while those in the right use a improved definition,
as described in the text.}
\end{figure}

The left column of plots in figure~3 presents the same comparison between different 
choices of external momentum in the definition of the mixing angle $\alpha$ but 
for a lower value of the pseudoscalar mass: $m_A=120$ GeV, with the rest of 
parameters chosen as in figure~2. The parameter $\tan\beta$ is 3 for the 
upper(-left) plot, 10 for the middle one and 40 for the lower plot. 
(In the plots for $\tan\beta=40$, curves with higher $p^2$ correspond
to lower values of $\tan[2\alpha(p^2)]$.) 
It is clear from this figure
that, although $p^2=p_*^2$ is somewhat better than other choices, there is 
some residual scale-dependence left. In this case with low $m_A$, the effect 
of higher order corrections is not negligible, and eventually such effects 
should be taken into account if a better scale stability of the mixing angle 
is required.

To do that, one should go beyond the one-loop approximation used so far. 
To identify more clearly the origin of the residual scale dependence let us
write explicitly [using (\ref{RGEm}), (\ref{RGEDm}) and (\ref{LL})] the elements 
of the radiatively-corrected mass matrix. Assume for simplicity that 
$\gamma_{ij}=\gamma_i \delta_{ij}$ and $m_{12}^2=m_{21}^2$. For the diagonal elements
one has (in one-loop leading-log approximation):
\be
M_{ii}^2(Q)\simeq m_{ii}^2(Q_0)-\Delta_{ii}(p,Q_0)-2\gamma_i(m_{ii}^2-p^2)
\ln{Q^2\over Q_0^2}\ ,
\label{diag}
\ee
and, for the off-diagonal element:
\be
M_{12}^2(Q)\simeq m_{12}^2(Q_0)-\Delta_{12}(p,Q_0)-(\gamma_1+\gamma_2)
m_{12}^2\ln{Q^2\over Q_0^2}\ .
\label{nondiag}
\ee
The scale at which the prefactors of the logarithmic terms
should be evaluated is irrelevant for the one-loop leading-log approximation:
different choices introduce differences only in higher order corrections. One 
may try to choose a scale that approximates well such corrections. One could
also argue in favour of including in the prefactors of the logarithmic terms
finite (non-logarithmic) radiative corrections. The key 
observation to improve the scale-independence of the mixing angle beyond one-loop
is the following: if the matrix elements have the form
\bea
M_{ii}^2(Q)&\simeq & {\tilde{m}}_{ii}^2-2\gamma_i({\tilde{m}}_{ii}^2-p^2)
\ln{Q^2\over Q_0^2}\ ,
\label{diagI}\\
M_{12}^2(Q)&\simeq &{\tilde{m}}_{12}^2(Q_0)-(\gamma_1+\gamma_2)
\tilde{m}_{12}^2\ln{Q^2\over Q_0^2}\ .
\label{nondiagI}
\eea
where ${\tilde{m}}_{ij}^2\equiv m_{ij}^2(Q_0) + [{\mathrm Q-indep.\ loop\ corrections }]$,
then it is straightforward to show that the mixing angle for the matrix with
elements (\ref{diagI}) and (\ref{nondiagI}) is exactly scale-independent for
$p^2=({\tilde{m}}_{11}^2+{\tilde{m}}_{22}^2)/2$. Therefore, in order to improve over 
the one-loop leading-log result, we make the replacement
\be
m_{ij}^2\rightarrow m_{ij}^2(Q_0)-\Delta_{ij}(p,Q_0)\ ,
\label{repl}
\ee
in the logarithmic terms of (\ref{diag}) and (\ref{nondiag}). 
At this point one should worry about the choice of scale $Q_0$. However, 
 eqs.~(\ref{RGEm}) and (\ref{RGEDm}) [with the replacement (\ref{repl}) made also in
(\ref{RGEm}) for consistency] guarantee that the elements
of the Higgs mass matrix, improved by (\ref{repl}), are independent of $Q_0$.
We have checked numerically that the impact of the choice of $Q_0$ in the mixing angle is tiny.

The replacement (\ref{repl}) is similar 
to the use of improved masses in \cite{andrea} for a numerically accurate 
calculation of radiatively corrected Higgs masses. It amounts to the inclusion
of some one-loop corrections to Higgs masses in the determination of $p_*^2$. Note however
that it does not correspond to the use of the full one-loop masses. Such choice would
be consistent only if the Higgs self-energies were computed at two-loops. If that is not
the case it does not give a better scale-independence than the advocated 
choice in (\ref{repl}). A consistent two-loop analysis of the scale-stability of mixing angles
would be interesting but lies beyond the scope of the present paper.

The results for $\tan[2\alpha(p^2)]$ after the improvement (\ref{repl}) are presented 
in the plots of the right column of figure~3. The parameters are the same as those
chosen for the left plots, so that the improvement in scale stability
can be appreciated by direct comparison: now the choice $p^2=p_*^2\equiv (M_{\tilt_1}^2+
M_{\tilt_2}^2)/2$ gives a perfectly scale-independent determination of $\alpha$.
From these plots we also see that, at least for moderate values of $\tan\beta$, 
there is a particular value of the renormalization scale for which a) the mixing 
angle is nearly momentum independent and all curves are focused in one point and 
b) the corresponding value of the mixing angle is a good approximation to the
scale independent result. Clearly, that choice of renormalization scale
corresponds to a value for which the one-loop logarithmic radiative
corrections are minimized. The existence of such a good choice of scale is
not always guaranteed if the spectrum of the particles in the loops is
widespread, case in which not all logarithms can be made small at a single
renormalization scale.

{\bf 4.} The mixing angle between two scalar particles with 
the same quantum numbers, once radiative corrections are taken into account,
depends on the renormalization scale and the external momentum. We have shown 
that, at one-loop, the scale dependence disappears for a particular choice of 
the external momentum, $p_*^2=(M_1^2+M_2^2)/2$, where $M_{1,2}$ are the masses
of the two particles. The particular momentum $p_*$ plays a role similar to the 
on-shell choice $p^2=M^2$ for the determination of a radiatively corrected 
physical mass $M$. 

We have applied this prescription to two cases of interest
in the Minimal Supersymmetric Standard Model: the mixing between stops and
the ${\cal CP}$-even Higgs mixing. We have shown numerically that the
advocated choice of momentum does indeed improve the scale independence of the
one-loop corrected mixing angles in both cases. In the Higgs boson case, especially
for low values of the pseudoscalar mass, we had to go beyond the one-loop 
approximation to get a satisfactory behaviour of the mixing angle, but this could
be achieved easily by taking into account higher order corrections (in particular,
one-loop non-logarithmic corrections to mass parameters in expression which were 
already of one-loop order). Therefore, our prescription could be very useful for 
a reliable determination of these mixing angles.

\section*{Acknowledgments}
We thank Abdel Djouadi, Jack Gunion, Michael Spira and especially Alberto Casas for
very helpful discussions.


\end{document}